\documentclass[twocolumn,amsmath,amssymb,aps,prl, superscriptaddress,floatfix,longbibliography]{revtex4-1}

\usepackage{hyperref}
\usepackage{graphicx}
\usepackage{mathtools}
\usepackage{multirow}
\usepackage{array}
\usepackage{tikz}
\usepackage{soul}

\begin{document}

\title{Loop-current order through the kagome looking glass}

\author{Rafael M. Fernandes}
\affiliation{Department of Physics, University of Illinois Urbana-Champaign, Urbana, Illinois 61801, USA}
\affiliation{Anthony J. Leggett Institute for Condensed Matter Theory, University of Illinois Urbana-Champaign, Urbana, Illinois 61801, USA}
\author{Turan Birol}
\affiliation{Department of Chemical Engineering and Materials Science, University of Minnesota, Minneapolis, Minnesota 55455, USA}
\author{Mengxing Ye}
\affiliation{Department of Physics and Astronomy, University of Utah, Salt Lake City, Utah 84112, USA}
\author{David Vanderbilt}
\affiliation{Department of Physics and Astronomy, Rutgers University, Piscataway, New Jersey 08854, USA}
\affiliation{Center for Materials Theory, Rutgers University, Piscataway, New Jersey 08854, USA}
\date{\today}

\begin{abstract}
In loop-current states, interacting electronic degrees of freedom collectively establish interatomic currents, in a rare example of magnetism in which spin degrees of freedom do not play the primary role. The main impact of such states on the electronic spectrum is not via the standard Zeeman term, but via the kinetic energy, in which hopping parameters develop non-trivial phases that break time-reversal symmetry. The recent proposal of loop-current states in kagome superconductors has stimulated renewed interest in this exotic type of magnetism. In this perspective, we use kagome materials as a scaffolding to frame the basic phenomenology of loop-current states. We provide an overview of the group-theoretical properties of loop currents, as well as of relevant microscopic models and \emph{ab initio} methods. Particular emphasis is given to the comparison with spin-density waves in the presence of spin-orbit coupling, as well as to the anharmonic coupling with charge-density waves, which is present in systems with threefold rotational symmetry. We also provide a brief overview of the current status of loop-current order in kagome metals and discuss open challenges including their experimental detection and interplay with other orders. 
\end{abstract}
\maketitle

\textbf{\emph{Introduction.}} Magnetism is a phenomenon most commonly associated with the spin degrees of freedom, and driven by exchange interactions. 
Magnetic order can have the same periodicity as the primitive cell, or a spin-density wave may arise at a nonzero wavevector $\bf Q$ that breaks the periodicity. In either case, the spin moments induce an orbital moment through the spin-orbit coupling (SOC). However, the orbital moments are generally weak and localized within an atomic sphere because of quenching by the crystal field. Thus, in almost all known cases, the instability that breaks time-reversal (TR) symmetry occurs in the spin sector.

However, it is also possible for a magnetic instability to
occur directly in the orbital sector. Because of the quenching, this is unlikely to mimic the on-site circulating currents associated with SOC-induced orbital moments. Instead, interest has focused on so-called ``loop current'' or
``flux'' states that are characterized by interatomic
currents flowing between atoms \cite{Affleck1991,Varma1997,Chakravarty2001,Sun2008}. In this case, the magnetism arises in the orbital sector, where
``orbital'' is simply a label for the spatial character
of the electronic wave functions. SOC can then induce
spin moments on atomic sites, but these are incidental to the loop-current instability. 

The potential mechanisms behind the loop currents
are a matter of active research.  In the absence of broken
TR symmetry, the electronic wave function $\psi({\bf r})$ can be taken as real, so that the local current density vanishes everywhere. A phase prefactor $e^{i\gamma({\bf r})}$ multiplied to this wave-function generates a current proportional to $\nabla\gamma$. This imposes a kinetic energy cost $\propto|\nabla\gamma|^2$ with no obvious energy
gain. As we shall see below, theoretical models of loop-current
states typically invoke specific mechanisms such as intersite or intervalley/interpatch interactions to explain this magnetic instability.

This type of electronic order can also lead to more exotic states of matter, such as an anomalous quantum Hall state \cite{Haldane1988}. The
question of whether loop-current magnetic states are realized in materials has been
discussed in diverse systems from cuprates \cite{Varma1997,Bourges2021} to colossal magnetoresistance materials \cite{Zhang2022}.
More recently, however, there 
has been a flurry of research activity in this front, motivated
partly by new experimental discoveries related to the $A$V$_{3}$Sb$_{5}$
kagome metals (with $A=\mathrm{Cs,\,K,\,Rb}$) \cite{Ortiz2019,Ortiz2021}, including a possible time-reversal symmetry-breaking (TRSB) charge order \cite{Mielke2022} that can also coexist with superconductivity.

The goal
of this focused perspective is to use the recent developments
related to kagome metals as a scaffolding to frame unresolved problems
and new challenges in loop-current (LC) magnetism. As
such, its focus will be on LC states that are intertwined with charge-density waves and break translational symmetry. We will
start with a phenomenological discussion, followed by microscopic
considerations, a discussion of \emph{ab initio }methods, and a brief
survey of the status of LC order in the $A$V$_{3}$Sb$_{5}$
kagome metals. We emphasize that it is beyond the scope of this paper
to review the rich literature on LC orders, including in
the context of high-$T_{c}$ superconductors \cite{Bourges2021}. Similarly, it is not the
goal of this perspective to review the field of $A$V$_{3}$Sb$_{5}$
kagome metals -- the interested reader is referred to Refs.\cite{Neupert2022,Wilson2024}. 

\textbf{\emph{Phenomenology}}. From a symmetry perspective, there
is no difference between a LC state that breaks translational
symmetry and a spin-density
wave (SDW) -- provided that spin-orbit coupling (SOC) is present \cite{Klug2018}.
This does not mean, however, that these
states are equivalent on the microscopic level. The SDW
order parameter is associated with the expectation value of the (staggered)
spin operator, whereas the LC order parameter corresponds to the expectation
value of the current operator (see Box 1 for details). Note, however, that a net non-dissipative
current is generally not allowed.

\begin{figure}
\centering
\includegraphics[width=0.95\columnwidth]{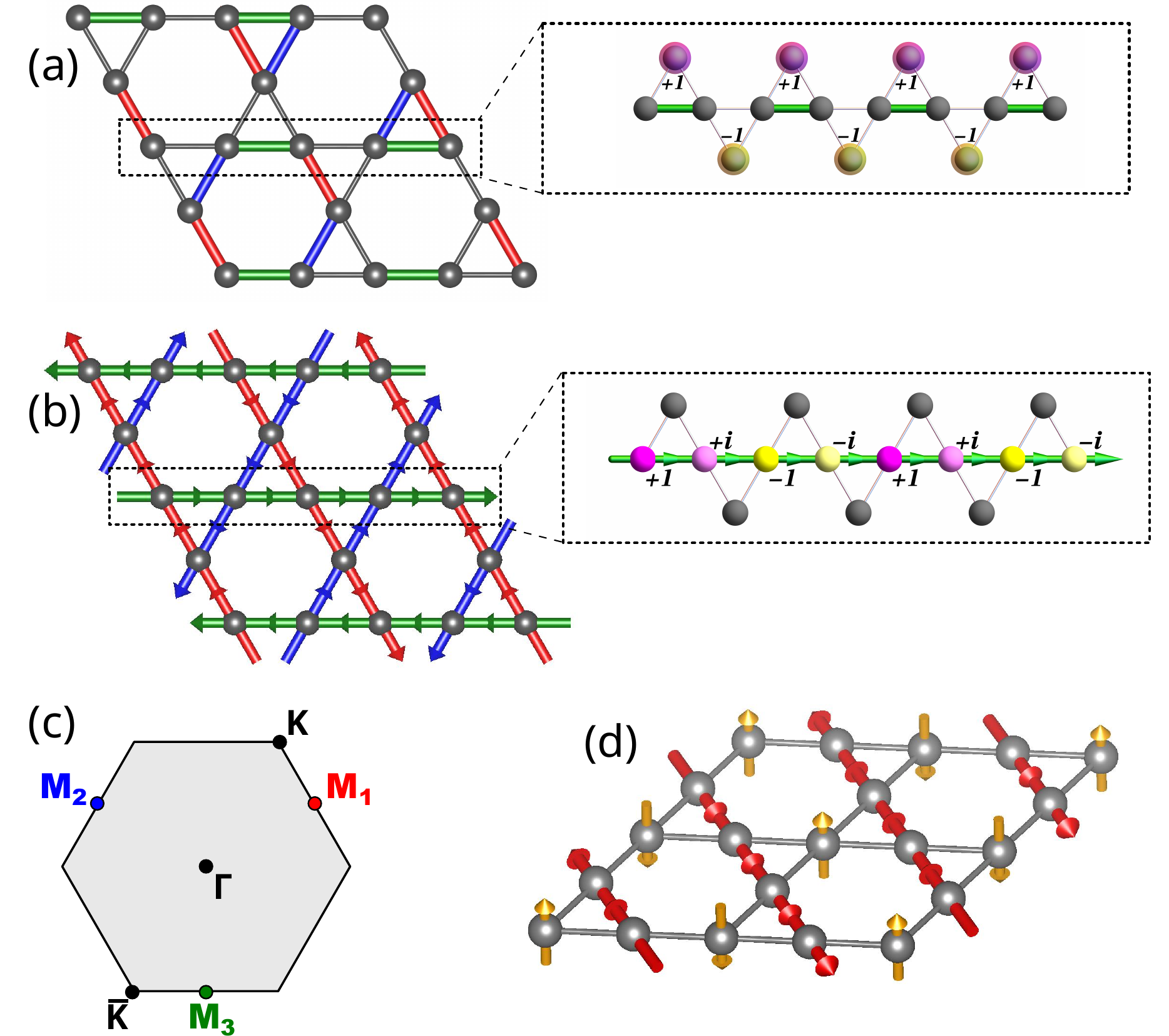}
\caption{\textbf{Loop current (LC), charge-density wave (CDW), and spin-density wave (SDW) orders}. Panels (a) and (b) illustrate, respectively, the CDW and LC orders on the kagome lattice formed by a linear combination of the three $\mathrm{M}$-point wave-vectors, which are shown in the Brillouin zone of panel (c), and given by $\mathrm{M}_{1}=\left(1/2,0,0\right)$,
$\mathrm{M}_{2}=\left(-1/2,1/2,0\right)$, and $\mathrm{M}_{3}=\left(0,-1/2,\,0\right)$. The colors of the bonds in (a) and of the currents in (b) correspond to the same-colored wave-vectors in panel (c). The insets of panels (a) and (b) show the phases $\exp(i \gamma)$ of the atomic wave functions, represented here by simple spherical orbitals, in the case of a single wavevector $\mathrm{M}_3$. In the CDW case, panel (a), the alternating phase $\gamma = 0,\pi$ corresponds to the site-CDW order (purple and yellow contours) that accompanies the bond-CDW order (green). In the LC case, the phases of the wave-functions of the atoms connected by the current change by $\pi/2$. Note that, because the point operations of the little group at $\mathrm{M}$ are isomorphic to the orthorhombic
point group $D_{2h}$, it admits eight one-dimensional irreps denoted by $M_{a}^{\pm}$ with $a=1,2,3,4$, with $\pm$ indicating parity under inversion. (d) LC order (red arrows) in the kagome lattice with a single wave-vector $\mathrm{M}_1$. An SDW order that has the same symmetry properties of the LC order is also shown as yellow arrows. Note that the spin configuration is non-uniform, since the spin magnetic moment vanishes on two of the sublattices. \label{fig:intro}}
\end{figure}

To further illustrate the differences between SDW and LC, consider their impact on the electronic states. The main effect of the SDW on the electronic Hamiltonian
is to create an inhomogeneous Zeeman term. In
contrast, LC order is primarily associated with the kinetic energy terms. Microscopically, the effect arises
from the coupling between the magnetic vector potential generated
by the loop currents and the electronic momentum (Box 1). As a result, in
a tight-binding description, LC order corresponds to a change in the phase of the hopping parameters $t_{ij}^{\mu\nu}$
between sites $i$ and $j$ and orbitals $\mu$ and $\nu$. Indeed, quite often, LC order will generate phases of
$\pm\pi/2$ in the hopping parameters involving neighboring sites.

To proceed, consider a system with negligible SOC,
so that SDW and LC orders can be formally treated as independent
instabilities. As discussed above, LC order modulates the \emph{phase}
of the hopping parameters, as illustrated in Fig. \ref{fig:intro}(b). In contrast, a charge-density wave (CDW)
modulates their \emph{amplitude} (Fig. \ref{fig:intro}(a)).
Note that we call CDW any
relative variation of the electronic charge regardless of its microscopic origin (i.e., Fermi surface nesting
or electron-phonon coupling). Also, while this charge variation can occur either on bonds (Su-Schrieffer-Heeger type) or on sites (Holstein type), as shown in Fig. \ref{fig:intro}(a), we hereafter focus only on the bond charge order for simplicity. 

The fact that the CDW and the LC orders modulate the phase and magnitude of the same quantity ($t_{ij}^{\mu\nu}$) highlights the close relationship between these two phenomena. Indeed, LC order that breaks translational
symmetry is often referred to as an ``imaginary'' CDW (denoted iCDW), as opposed to a traditional ``real'' CDW (denoted rCDW) \cite{Chubukov2009}. This
nomenclature, while useful, should not be interpreted as a statement that
the imaginary component of a CDW order parameter is equivalent to
LC. Group theory provides the proper language
to make these distinctions sharper. Hereafter, we
will denote the CDW order parameter as $W$ and the LC order parameter
as $\Phi$. For a given lattice and wave-vector $\mathbf{Q}$, the
order parameters can be classified according to the irreducible representations
(irreps) of the space group at $\mathbf{Q}$. Because we focus on LC order emerging from a paramagnetic phase, $\Phi$ can be classified
as a time-reversal symmetry-breaking (TRSB) irreducible representation of the space group. We use a lower case $m$ to denote that the irrep is odd under time-reversal. Thus, for a given time-reversal even irrep $\Gamma$ of the space group, the notation  $m\Gamma$ denotes the corresponding irrep that is odd under time-reversal. 

\begin{table*}
\begin{centering}
\begin{tabular}{|c|c|c|c|c|c|}
\hline 
LC & CDW & magnetism & AHE / Kerr & resistivity anisotropy & piezomagnetism\tabularnewline
\hline 
\hline 
$\left(\Phi_0,\Phi_0,\Phi_0\right)$ & $\left(W_0,W_0,W_0\right)$ & ferromagnetic & \textcolor{green}{$\checkmark$} & \textcolor{red}{$\times$} & \textcolor{red}{$\times$} \tabularnewline
\hline 
$\left(\Phi_0,\Phi_0,0\right)$ & $\left(0,0,W_0\right)$ & antiferromagnetic & \textcolor{red}{$\times$}  &  \textcolor{green}{$\checkmark$} & \textcolor{red}{$\times$} \tabularnewline
\hline 
$\left(\Phi_0,0,-\Phi_0\right)$ & $\left(W_0,\tilde{W}_0,W_0\right)$ & ferro-octupolar & \textcolor{red}{$\times$}  &  \textcolor{green}{$\checkmark$} &  \textcolor{green}{$\checkmark$}\tabularnewline
\hline 
\end{tabular}
\par\end{centering}
\caption{Common LC-CDW phases formed by different combinations of the loop-current
(LC) order parameter $\Phi=\left(\Phi_{1},\Phi_{2},\Phi_{3}\right)$
and the charge-density wave (CDW) order parameter $W=\left(W_{1},W_{2},W_{3}\right)$, as illustrated in Fig. \ref{fig:phases}. 
The third column shows the magnetic properties of each phase, whereas
the last three columns show their experimental manifestations in terms
of the anomalous Hall effect (AHE), spontaneous Kerr effect, threefold
rotational symmetry-breaking (resistivity anisotropy)
and piezomagnetism. \label{tab:classification}}
\end{table*}

The irreps at momentum $\mathbf{Q}$ are formed by irreps of the little group (crystallographic operations that leave the momentum unchanged) and an irrep that encodes how $\mathbf{Q}$ is connected to its symmetry-equivalent momenta (known as the star of $\mathbf{Q}$). For concreteness, consider the case of a simple orthorhombic space group, and a two-dimensional irrep at $\mathbf{Q}$ with only two vectors in its star, $\mathbf{Q}=\left(\mp Q_{x},0,0\right)$. The CDW order parameter $W$ must have two
components, $W=\left(W_{1},\,W_{2}\right)$, which can be interpreted
in terms of the Fourier components of the charge density, $\rho_{\mathbf{Q}}$.
Since $\rho_{\mathbf{Q}}^{*}=\rho_{-\mathbf{Q}}$, one can associate
$W=\left(W_{1},\,W_{2}\right)$ to either $\left(\rho_{\mathbf{Q}},\rho_{-\mathbf{Q}}\right)$,
in which case the irreps are complex-valued, or to $\left(\mathrm{Re}\left(\rho_{\mathbf{Q}}\right),\,\mathrm{Im}\left(\rho_{\mathbf{Q}}\right)\right)$,
in which case the irreps are real-valued. The latter convention makes
it transparent that, regardless of whether the CDW order parameter
is purely imaginary or purely real, it still describes a TR
preserving charge order (i.e. a ``real'' CDW). The main difference
between the configurations $W=\left(W_{0},0\right)$, in which the
CDW order parameter is purely real, and $W=(0,W_{0})$, where the
CDW order parameter is purely imaginary, is that the former is even under
inversion symmetry whereas the latter is odd; both nevertheless
preserve time-reversal symmetry. We also underline the fact that LC order is
not synonymous with chiral CDW order. Chiral order means that the system lacks
any improper rotations (including mirrors and inversion centers) as symmetry elements, which is different from lacking time-reversal symmetry. 

The key point is that one cannot associate the LC order parameter
$\Phi$ to the imaginary part of the CDW order parameter $W$. This
becomes even clearer for time-reversal invariant momenta ($-\mathbf{Q}=\mathbf{Q}$). Since this is the case for the $A$V$_{3}$Sb$_{5}$ kagome metals, it is instructive to
discuss it in more depth \cite{Christensen2021,Christensen2022,Fischer2023}. The space group of these materials is P6/mmm,
which is the ``most symmetric'' hexagonal space group. Consider a CDW wave-vector located at the $\mathrm{M}$
point of the Brillouin zone, illustrated in Fig.\ref{fig:intro}(c) (in these materials,
the CDW wave-vector is actually at the $\mathrm{L}$ point, which
is displaced from the $\mathrm{M}$ point along the $q_{z}$ axis
by $Q_{z}=1/2$). Because there are three $\mathrm{M}_i$ points related by three-fold rotations, the CDW order parameter has three components, $W=\left(W_{1},\,W_{2},\,W_{3}\right)$, with each $W_{i}$ associated with the Fourier component of the charge density at the corresponding wave-vector $\mathrm{M}_{i}$. Since $\mathrm{M}_i$ are time-reversal invariant-momenta, $W_{i}$ are real-valued. Moreover, as explained in the caption of Fig. \ref{fig:intro}, group theory says that these three-component order parameters must transform as one of eight space-group irreps denoted $M_{a}^{\pm}$ (with $a=1,2,3,4$). The triple-$\mathbf{Q}$ configuration of CDW bond order $W=(W_0,W_0,W_0)$, as well as a CDW site order transforming as the same irrep $M_1^+$, are shown in Fig. \ref{fig:intro}(a).

There are also eight possible three-component LC order
parameters $\Phi=\left(\Phi_{1},\,\Phi_{2},\,\Phi_{3}\right)$, which transform as the irreps $mM_{i}^{\pm}$ and break TR symmetry. The triple-\textbf{Q
} LC configuration with $\Phi=(\phi_0,\phi_0,\phi_0)$ transforming as the irrep  $mM_2^+$ is shown in Fig. \ref{fig:intro}(b). In Fig. \ref{fig:intro}(d), a single-\textbf{Q
}loop-current configuration is shown together with an SDW with the same symmetry properties. Note that different loop-current configurations may transform as the same irrep \cite{JPHu2021}.
Thus, while it
can be theoretically insightful to associate $\Phi$ and $W$ as real and imaginary parts
of a composite complex order parameter that transforms as a reducible representation \cite{Lin2021,Park2021}, it is important
to keep in mind that a hypothetical degeneracy between the two components is necessarily lifted by the crystal symmetries ~\cite{Park2021}. 

A crucial aspect of the problem is that the three-fold rotational symmetry of the kagome lattice allows anharmonic couplings between the CDW and LC order parameters, manifested as third-order terms in the coupled free energy of $W$ and $\Phi$ \cite{Park2021,Christensen2022}. The main consequence of this coupling is that the condensation of an LC order parameter with components at two distinct wave-vectors $\mathrm{M}_{i_1}$ and $\mathrm{M}_{i_2}$ must trigger a CDW order-parameter component at the third wave-vector $\mathrm{M}_{i_3} = \mathrm{M}_{i_2} + \mathrm{M}_{i_2}$, i.e., $W_{i_1} \sim \Phi_{i_2} \Phi_{i_3}$ (with $i_1$, $i_2$, and $i_3$ all different from each other). In
other words, a modulation of the phases of the hopping parameters typically induces a modulation in their amplitudes. 

\begin{figure}
\centering
\includegraphics[width=0.85\columnwidth]{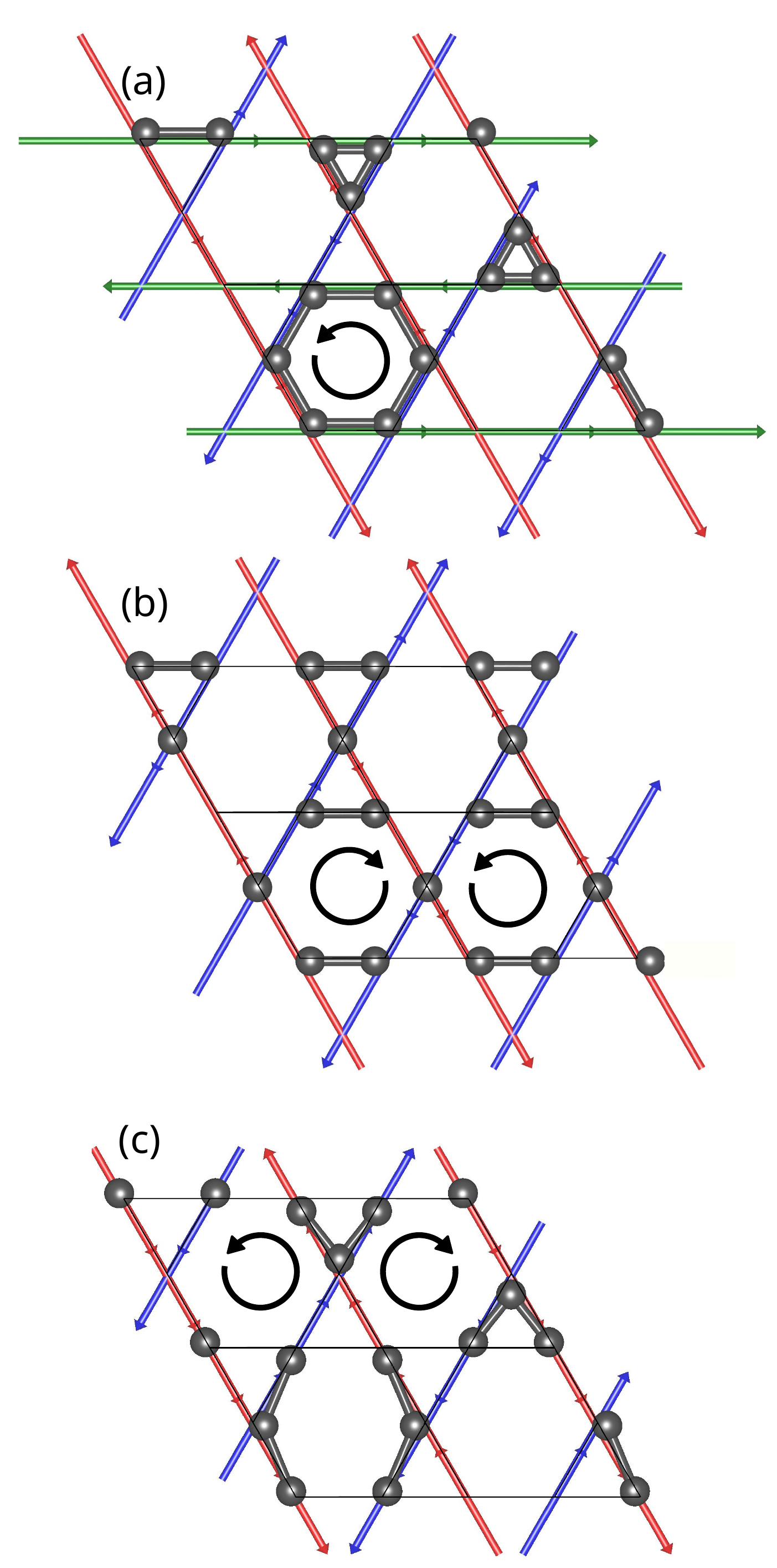}
\caption{\textbf{Different types of coupled LC and CDW orders on the kagome lattice}. Because of the anharmonic coupling between the two order parameters, LC order is in most cases accompanied by CDW order. Each panel illustrates one of the three coupled LC-CDW phases discussed in the main text and in Table \ref{tab:classification}. Panel (a): 3\textbf{Q}-3\textbf{Q} configuration $\Phi=\left(\Phi_{0},\Phi_{0},\Phi_{0}\right)$ and $W=\left(W_{0},W_{0},W_{0}\right)$. Panel (b): 2\textbf{Q}-1\textbf{Q} configuration $\Phi=\left(\Phi_{0},\Phi_{0},0\right)$ and $W=\left(0,0,W_{0}\right)$. Panel (c): 2\textbf{Q}-3\textbf{Q} configuration $\Phi=\left(\Phi_{0},0,-\Phi_{0}\right)$ and $W=\left(W_{0},\tilde{W}_{0},W_{0}\right)$.\label{fig:phases}}
\end{figure}

As a result of this non-trivial coupling, the phase diagram of the $\mathrm{M}$-point
coupled LC-CDW orders on hexagonal crystals can be very
rich. While the many independent coefficients of the Landau free energy render it challenging
to make generic statements about the stable phases, a few flux-CDW states stand out \cite{Christensen2022,Park2021,Fischer2023}, as shown in Table \ref{tab:classification} and Fig. \ref{fig:phases}. The only phase
that allows for LC order and no CDW order is the single-\textbf{Q
}LC configuration $\Phi=\left(\Phi_{0},0,0\right)$ (and, of course, its symmetry-equivalent domains). In contrast, the multi-\textbf{Q
}LC phases will necessarily be accompanied by CDW order.

The magnetic properties of these mixed LC-CDW phases are also very rich. For concreteness,
consider the simplest types of combined CDW and LC order as shown in Fig.~\ref{fig:phases}, which involve combinations of $M_{1}^{+}$ CDW and $mM_{2}^{+}$ LC irreps. For the
3\textbf{Q}-3\textbf{Q} configuration described by $\Phi=\left(\Phi_{0},\Phi_{0},\Phi_{0}\right)$
and $W=\left(W_{0},W_{0},W_{0}\right)$, shown in Fig. \ref{fig:phases}(a) (top row of Table I), the magnetic moments generated
by the loop currents do not cancel each other exactly, resulting in
a net ferromagnetic moment. The anharmonic coupling between the LC and CDW order parameters
is essential for this result, as the bond distortions generated by
the loop currents make the local environment experienced by the currents
inequivalent \cite{Christensen2022}. On the other hand, as shown in Fig.~\ref{fig:phases}(b) (middle row of Table I), in the 2\textbf{Q}-1\textbf{Q} configuration
described by $\Phi=\left(\Phi_{0},\Phi_{0},0\right)$ and $W=\left(0,0,W_{0}\right)$,
the magnetic moments generated by the loop currents cancel. This configuration has the symmetry of an
antiferromagnet, in that it possess magnetic translation symmetry.
In contrast to the 3\textbf{Q}-3\textbf{Q} state, however, it breaks
the six-fold rotational symmetry of the kagome lattice. Another interesting phase, displayed in Fig. \ref{fig:phases}(c) (bottom row of Table \ref{tab:classification}), is the 2\textbf{Q}-3\textbf{Q} state
described by $\Phi=\left(\Phi_{0},0,-\Phi_{0}\right)$ and $W=\left(W_{0},\tilde{W}_{0},W_{0}\right)$,
dubbed the congruent CDW flux phase \cite{Xing2024}. In this case, the loop-current fluxes cancel,
implying a vanishing net magnetic dipole moment, but the system still
has a net magnetic octupole moment, which gives rise to piezomagnetism.

\textbf{\emph{Microscopic Mechanisms}}. The group-theory analysis
provides important insights into the properties of the LC order, but
it cannot establish which microscopic mechanisms promote this electronic
instability. This requires either low-energy microscopic models of interacting
electrons, or \emph{ab initio }calculations on
materials.

\begin{figure}
\centering
\includegraphics[width=0.95\columnwidth]{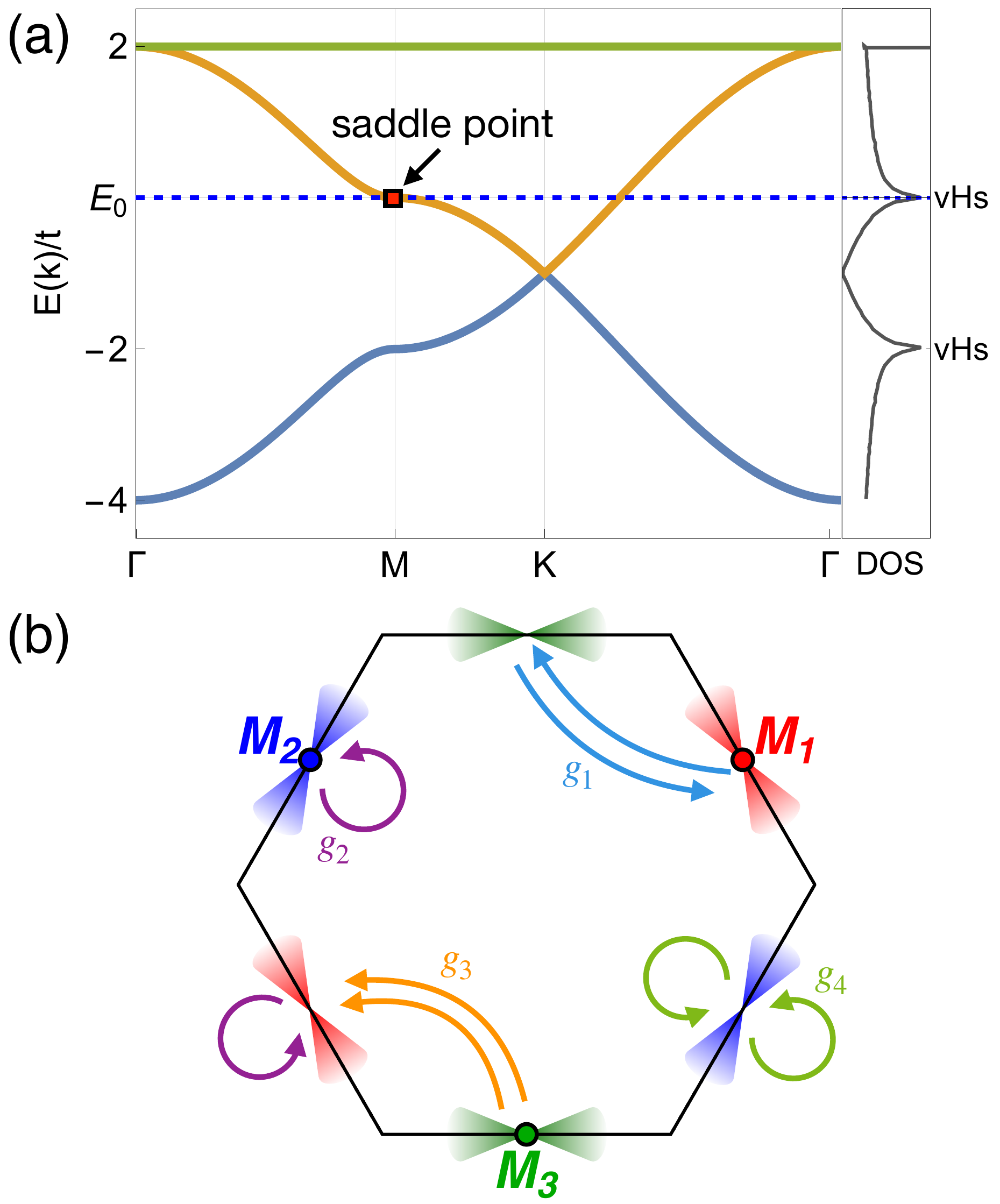}
\caption{\textbf{Saddle points, van Hove singularities, and the patch model}. Panel (a) shows the band structure of the nearest-neighbor tight-binding model on the kagome lattice, highlighting the saddle points at the $\mathrm{M}$ point, which display a minimum along $\Gamma$-$M$-$\Gamma$ but a maximum along $K$-$M$-$\bar{K}$. The saddle points give rise to a logarithmically diverging density of states (DOS) at the corresponding chemical potential. Panel (b) illustrates the patch model, highlighting the four different types of interactions coupling the electronic states around each $M$-point van Hove singularity \cite{Park2021}. \label{fig:microscopics}}
\end{figure}

Different models have been proposed to capture the $\mathrm{M}$-point LC order relevant to the kagome metals, most of which have focused on the interplay between van Hove singularities, longer-range Coulomb repulsion, and the multi-orbital or multi-sublattice character of these systems ~\cite{Denner2021,Park2021,Tazai2022,Ferrari2022,Ziqiang2023,Scammell2023,Wu2023,Chen2024flux,XianxinWu2024,HYKee2024}. 
As an illustration, we consider the patch model, which focuses on the low-energy electronic states
around the $\mathrm{M}$ points~\cite{Lin2021,Park2021}. The motivation for this model is that simple tight-binding
models on hexagonal lattices \cite{Lin2021} exhibit electronic dispersions $E(\mathbf{k})$ with saddle
points at the $\mathrm{M}$ points, as shown in Fig. \ref{fig:microscopics}(a) for the kagome lattice
\begin{equation}
E\left(\mathbf{k}+\mathbf{Q}_{\mathrm{M}}\right)\approx E_{0}+\frac{\hbar^{2}\left(k_{x}^{2}-k_{y}^{2}\right)}{2m^{*}} , \label{eq:vHs}
\end{equation}
where $k_{x}$ and $k_y$ point along the $\Gamma$-$\mathrm{M}$ and $\mathrm{M}$-$\mathrm{K}$ directions, respectively,
and $m^{*}$ is the effective mass. In two dimensions, the density
of states (DOS) displays a logarithmic
divergence at $E_{0}$ 
(Fig. \ref{fig:microscopics}(a)), whereas in three dimensions it displays a
kink. These are known as van Hove singularities (vHs), which dominate the low-energy behavior if the chemical potential is close to $E_{0}$. 

The patch model defines three ``flavors''
of electronic states, one for each $\mathrm{M}$ point, as well as four symmetry
allowed interactions $g_i$ between them, as seen in Fig.~\ref{fig:microscopics}(b). Because of the divergent DOS, weak-coupling approaches, such as the renormalization group, can
be employed. The results (details in Box 2) reveal that, depending on which of the four interactions are dominant, the resulting instability of the model is towards a LC state.   

Since most microscopic models rely on vHs, it is important to discuss the band structure of $A$V$_{3}$Sb$_{5}$. In these materials, the vHs originate
from the V $d$-orbitals \cite{Neupert2022,Kang2022}. The saddle points are close to the Fermi level on the $q_z=0$ plane, and there is sufficient dispersion along $q_z$ that they cross the Fermi level at non-zero $q_{z}$ \cite{Jeong2022,Ritz2023}. Importantly, there are at least four vHs with comparable
energies but different symmetries \cite{Christensen2022}. This allows extensions of the simple patch model to ``multi-flavor''
patch models, which reveal interesting new physics \cite{Scammell2023,HYKee2024}.

Beyond low-energy models, there is a surprising dearth of \emph{ab initio} studies on LCs, most probably due to the inability of commonly employed local density or generalized gradient approximations (LDA or GGA) to reproduce the LC state. In these approximations, the exchange-correlation functional is spatially local, which makes them unlikely to stabilize a nonzero expectation value of the favorable interactions $g_j$ in Fig.~\ref{fig:microscopics}(c), which have significant intersite character. Nevertheless, the \emph{ab initio} calculation in Ref.~\cite{Liu2016} predicted a LC phase in the absence of spin ordering in CrSi(Ge)Te$_3$-based systems. It may be significant that this work employed a hybrid functional \cite{Heyd2003hybrid}, which introduces a fraction of nonlocal Fock exchange to the exchange-correlation energy.
 
\textbf{\emph{Experimental Status}} \textbf{\emph{and Outlook}}. Whether
and what type of LC order is realized in the $A$V$_{3}$Sb$_{5}$ kagome metals remain a widely investigated issue. These materials display
a CDW transition temperature on the order of $100$ K, followed by
a superconducting transition at much lower temperatures, on the order
of $1$ K \cite{Wilson2024}. One of the pieces of evidence
in favor of TRSB inside the CDW state is muon spin rotation ($\mu$SR)
measurements. They reveal an enhancement of the muon relaxation
rate below the CDW transition, followed by another enhancement at
a lower temperature, which are attributed to internal magnetic
fields of $\sim0.01$~mT arising from the electronic degrees
of freedom \cite{Mielke2022,Yu2021flux,Khasanov2022,Guguchia2022,Bonfa2024}. There are also reports of a pronounced response of the
CDW state to a magnetic field, which is again consistent with
LC order. STM measurements in all three members of this kagome family
find that the relative amplitude between the CDW Bragg peaks changes
sign depending on the direction of the out-of-plane applied magnetic
field \cite{Jiang2021,Shumiya2021,Deng2024,Xing2024}; Ref. \cite{Candelora2024}, however, does not report such a behavior. Ref. \cite{Jiang2021} interpreted
this response in terms of a chiral CDW, whereas Ref. \cite{Xing2024} attributed this
behavior to piezomagnetism, i.e., a shear strain triggered by the magnetic
field in linear order. Besides STM, transport measurements find that
an in-plane resistivity anisotropy emerges inside the CDW state only
when a magnetic field is applied \cite{Guo2024}, a behavior that was also associated
with the condensation of loop-current order \cite{Fischer2023}. Further evidence for TRSB
come from nonlinear transport measurements that found a magnetochiral
anisotropy \cite{Guo2022}, which also requires broken mirror symmetries, as well as from torque magnetometry data \cite{Matsuda2024}.

Direct measurements of a magnetic moment remain controversial. A ``zero-field''
anomalous Hall effect, in which the Hall conductivity retains a non-zero
value after the applied field is removed, has not been reported. Nevertheless,
large responses of the Hall conductivity \cite{Yu_AHE2021} and of the Nernst coefficient \cite{Felser2022}
to an applied magnetic field were found, whose origins remain unsettled \cite{Welp_AHE2024}. Moreover, spin-polarized
STM did not find evidence for magnetic moments \cite{HHWen2022}. A spontaneous Kerr
effect, which requires non-zero magnetization (i.e. ferromagnetism), was reported in Refs. \cite{Xu2022Three-state,Hu2022time}
below the CDW transition. However, Refs. \cite{Saykin2023Kerr,Farhang2023} did not observe a spontaneous
Kerr effect, with the former noting instead that a strong diamagnetic contribution
to the susceptibility appears below the CDW transition. A spin-polarized
neutron diffraction study observed only a hint of weak magnetic scattering
at the corner of the second Brillouin zone, close to the detection
limit, but consistent with a moment of the order of $0.01\mu_{B}$
per vanadium triangle \cite{Liege2024}. 

In interpreting these experimental results, it is important to emphasize
that an LC state does not necessarily induce a net magnetization,
since TRSB is not synonymous with ferromagnetism. This implies that the
absence of effects that require the presence of a net magnetic dipole
moment, such as the spontaneous Kerr effect and the anomalous Hall
effect, is not necessarily inconsistent with a LC state. Indeed, in
the flux CDW states discussed in Table \ref{tab:classification}, only the 3\textbf{Q}-3\textbf{Q}
state displays a nonzero magnetization, whereas the loop-current patterns
of the 2\textbf{Q}-1\textbf{Q} and 2\textbf{Q}-3\textbf{Q} states
break time-reversal symmetry in different ways \cite{Christensen2022}. For instance, the
rotational-symmetry-breaking 2\textbf{Q}-1\textbf{Q} state has no
net magnetic moments, displaying instead a flux configuration analogous
to a spin antiferromagnetic phase. The 2\textbf{Q}-3\textbf{Q} congruent
CDW flux phase, on the other hand, displays a net magnetic octupole
moment, which is manifested experimentally as piezomagnetism \cite{Xing2024}. As pointed
out in Ref. \cite{Xing2024}, if indeed the LC state is piezeomagnetic, residual strain
can locally condense a nonzero magnetization, which could be picked
up, for instance, by Kerr-effect measurements. Moreover, while the anharmonic coupling between LC and CDW implies that LC order necessarily triggers CDW
order, the converse is not necessarily true; nevertheless, this coupling
can favor the emergence of LC order inside the CDW phase \cite{Christensen2021}. Overall, it is clear that the CDW phase of the $A$V$_{3}$Sb$_{5}$
kagome metals displays sizable responses to magnetic fields and strain,
as discussed in Ref. \cite{Guo2024}. While this makes the identification of the
potential LC state challenging, it also opens the door for new physical
behaviors. 

In summary, the possible realization of loop-currents in kagome metals provides new opportunities to elucidate and investigate this unusual type of magnetism. These include new experimental setups to probe orbital magnetism that do not rely solely on magnetization measurements, such as piezomagnetism or magneto-electric effects.  On the theory front, the open challenges encompass \emph{ab initio} predictions of loop-current order, as well as understanding its interplay with other electronic degrees of freedom. In particular, the interplay with charge order, which occurs naturally in systems with threefold symmetry, is a central aspect of this state's phenomenology. Moreover, the interplay with superconductivity, which is realized in kagome metals at temperatures much lower than the charge-order one, remains an open question. It has been recently shown that intra-unit-cell loop-current fluctuations associated with broken inversion symmetry are detrimental to pairing \cite{Palle2024}. However, it remains little explored the scenario in which loop-current fluctuations at finite wave-vectors, like those discussed in this perspective, mediate superconductivity.

\begin{acknowledgments}
We thank B. Andersen, L. Balents, M. Christensen, V. Madhavan, T. Park, E. Ritz, H. Roising, Z. Wang, S. Wilson and I. Zeljkovic for fruitful discussions. R.M.F. was supported by the Air Force Office of Scientific Research under Award No. FA9550-21-1-0423. D.V. was supported by NSF Grant DMR-2421845. T.B. was supported by the NSF CAREER grant DMR2046020.
\end{acknowledgments}

\appendix

\section{Box 1: Microscopic definitions of the LC and SDW order parameters}

Consider a simple model of electrons hopping on the sites of a lattice.
We introduce the $N$-dimensional spinor $\Psi_{\mathbf{k},s}=\left(\psi_{\mathbf{k},s}^{(\mu_{1})},\cdots,\psi_{\mathbf{k},s}^{(\mu_{N})}\right)^{T}$,
where $\mu_{i}$ denotes an internal degree of freedom like orbital
or sublattice, such that $\psi_{\mathbf{k},s}^{(\mu_{i})}$ annihilates
an electron at orbital $\mu_{i}$ with momentum $\mathbf{k}$ and
spin projection $s$. SDW order with wave-vector $\mathbf{Q}$ corresponds
to a non-zero expectation value of the spin operator,
\begin{equation}
\boldsymbol{\Phi}_{\mathrm{SDW}}=\frac{1}{N_s}\sum_{\mathbf{k},ss'}\left\langle \Psi_{\mathbf{k},s}^{\dagger}\boldsymbol{\sigma}_{ss'}\Psi_{\mathbf{k}+\mathbf{Q},s'}^{\phantom{\dagger}}\right\rangle ,
\end{equation}
where $\boldsymbol{\sigma}$ is a vector of Pauli matrices and $N_s$ (not to be confused with the number of internal degrees of freedom $N$)
is the number of sites. Thus, in general $\boldsymbol{\Phi}_{\mathrm{SDW}}$
is a vector in spin space (by virtue of the vector of Pauli matrices $\boldsymbol{\sigma}$) and a matrix in orbital space (by virtue of the $N$-dimensional spinor $\Psi_{\mathbf{k},s}=\left(\psi_{\mathbf{k},s}^{(\mu_{1})},\cdots,\psi_{\mathbf{k},s}^{(\mu_{N})}\right)^{T}$). By construction,
it modifies the electronic Hamiltonian via a Zeeman term
\begin{equation}
H_\textrm{Z} = -g \sum_{\mathbf{k},ss'} \boldsymbol{\Phi}_{\mathrm{SDW}} \cdot \Psi_{\mathbf{k},s}^{\dagger}\boldsymbol{\sigma}_{ss'}\Psi_{\mathbf{k}+\mathbf{Q},s'}^{\phantom{\dagger}} ,
\end{equation}
where $g$ is an appropriate coupling constant. In contrast, the case of LC order with wavevector $\mathbf{Q}$ corresponds
to a non-zero expectation value of a combination of current operators.
In momentum space, it can be cast in the general form
\begin{equation}
\Phi_{\mathrm{LC}}=\frac{1}{N}\sum_{\mathbf{k},s}\left\langle \Psi_{\mathbf{k},s}^{\dagger}\gamma_{\mathbf{k},\mathbf{k}+\mathbf{Q}}\Psi_{\mathbf{k}+\mathbf{Q},s}^{\phantom{\dagger}}\right\rangle 
\end{equation}
where $\gamma_{\mathbf{p},\mathbf{q}}$ is a matrix in the $N$-dimensional orbital space with matrix elements $\gamma_{\mathbf{p},\mathbf{q}}^{(\mu_i, \mu_j)}$,
that must satisfy the constrains $\gamma_{\mathbf{q}_{1},\mathbf{q}_{2}}^{\dagger}=-\gamma_{\mathbf{q}_{2},\mathbf{q}_{1}}=\gamma_{-\mathbf{q}_{1},-\mathbf{q}_{2}}^{T}$
(assuming that time reversal does not change the internal index $\mu_{i}$).
Importantly, $\Phi_{\mathrm{LC}}$ modifies the Hamiltonian via the minimal substitution
in the kinetic energy term. In a tight-binding description, it
corresponds to the so-called Peierls substitution, in which the hopping
parameter $t_{ij}$ between sites $i$ and $j$ acquires a phase
\begin{equation}
t_{ij}\rightarrow t_{ij}\exp\left(-\frac{ie}{\hbar c}\int_{i}^{j}\mathbf{A}\left(\mathbf{r}\right)\cdot d\mathbf{r}\right) ,
\end{equation}
where $\mathbf{A}\left(\mathbf{r}\right)$ is the vector potential.
In the case of loop currents, the relevant quantity is the flux generated
by the closed loops.

\section{Box 2: The patch model}

The patch model (see, e.g., Ref. \cite{Park2021}) focuses on the low-energy electronic state near the van Hove singularities at the $\mathrm{M}$ points, which can be expressed in terms of electron annihilation operators $\psi_{\mathbf{M}_i,s}$. An order parameter in the charge channel at momentum $\mathbf{M}_i = \mathbf{M}_l-\mathbf{M}_j$ for this simple case can be written as (here, $i,j,k$ are non-repeating indices)
\begin{equation}
 \mathcal{O}^{\pm}_{\mathbf{M}_i} = \frac{1}{N}\sum_{\mathbf{k},s} \langle \psi^{\dagger}_{\mathbf{M}_j + \mathbf{k},s}\psi^{\phantom{\dagger}}_{\mathbf{M}_l + \mathbf{k},s} \pm \psi^{\dagger}_{\mathbf{M}_l + \mathbf{k},s}\psi^{\phantom{\dagger}}_{\mathbf{M}_j + \mathbf{k},s}\rangle .
\end{equation}
The plus sign gives a real order parameter, which corresponds to the bond charge order parameter (rCDW) of the main text, $\mathcal{O}^{+} \rightarrow W$. On the other hand, the minus sign gives an imaginary order parameter corresponding to the loop-current order parameter (iCDW) of the main text, $\mathcal{O}^{-} \rightarrow -i \Phi$. We emphasize that the rCDW and iCDW order parameters transform differently under time reversal, and in general have different transition temperatures. 

There are four distinct types of interactions $g_i$ involving the patch fermionic operators $\psi_{\mathbf{M}_i,s}$. Interactions $g_{1}$, $g_{2}$, and $g_{3}$ are the inter-patch exchange,
density-density, and Umklapp interactions, respectively, whereas $g_{4}$ is the intra-patch density-density interaction, as seen in Fig.~\ref{fig:microscopics} (b). In the absence of SOC, there are four possible distinct density-wave instabilities at $\mathbf{M}_{i}$, corresponding to ``real'' and ``imaginary'' charge-density
waves and ``real'' and ``imaginary'' spin-density
waves (rSDW and iSDW). The iSDW corresponds to a state with spin-current loops. In the presence of SOC, iCDW and rSDW are symmetry-equivalent, as are the iSDW and rCDW.

The leading instability is determined by the inter-patch interactions: $g_{1}>0$ favors an instability in the spin channel, whereas $g_{1}<0$ favors one in the charge channel. $g_{2}>0$, on the other hand, favors both channels equally. Finally, $g_{3}>0$ favors an instability in the iCDW (or rSDW) channel over rCDW (or iSDW) channel. Thus, the iCDW phase is expected when $g_{1}<0$, $g_{2}>0$, and $g_{3}>0$. We note that these interactions are ``flowing interactions'' in the renormalization group sense, in that they are renomalized due to the interference between different channels. Indeed, besides these four density-wave channels, the system can also be unstable towards a superconducting instability either in the conventional $s$-wave channel or in the unconventional $d+id$ channel \cite{Lin2021,Park2021}.

\bibliography{loop_currents}

\end{document}